\documentclass[aps,prb,superscriptaddress,twocolumn]{revtex4-1}

\usepackage{lineno,hyperref}
\usepackage{graphicx}
\usepackage{bmpsize}
\usepackage{amsfonts}

\newcommand{\ben}{\begin{eqnarray}}
\newcommand{\een}{\end{eqnarray}}

\newcommand{\bef}{\begin{figure}[h!bt]\centering}
\newcommand{\eef}{\end{figure}}
\newcommand{\bet}{\begin{table}[hbt]\centering}
\newcommand{\eet}{\end{table}}
\begin{document}

\title{Significant change in the electronic behavior associated with structural distortions in the single crystalline SrAg$_4$As$_2$}

\author{Bing Shen}
\affiliation{Department of Physics and Astronomy and California NanoSystems Institute, University of California, Los Angeles,
CA 90095, USA}
\author{Eve Emmanouilidou}
\affiliation{Department of Physics and Astronomy and California NanoSystems Institute, University of California, Los Angeles,
CA 90095, USA}

\author{Xiaoyu Deng}
\affiliation{Department of Physics and Astronomy, Rutgers University, Piscataway, NJ 08854, USA}
\author{Alix McCollam}
\affiliation{High Field Magnet Laboratory (HFML-EMFL), Radboud University, 6525 ED Nijmegen, The Netherlands}
\author{Jie Xing}
\affiliation{Department of Physics and Astronomy and California NanoSystems Institute, University of California, Los Angeles, CA 90095, USA}

\author{Gabriel Kotliar}
\affiliation{Department of Physics and Astronomy, Rutgers University, Piscataway, NJ 08854, USA}

\author{Amalia I. Coldea}
\affiliation{Clarendon Laboratory, Department of Physics, University of Oxford, Parks Road, Oxford OX1 3PU, UK}
\author{Ni Ni}
\email{Corresponding author: nini@physics.ucla.edu}
\affiliation {Department of Physics and Astronomy and California NanoSystems Institute, University of California, Los Angeles, CA 90095, USA}

\begin{abstract}

Here we report a combined study of transport and thermodynamic measurements on the layered pnictide material SrAg$_4$As$_2$. Upon cooling, a drop in electrical and Hall resistivity, a jump in heat capacity and an increase in susceptibility and magnetoresistance (MR) are observed around 110 K. All suggest non-magnetic phase transitions emerge at around 110 K, likely associated with structural distortions. In sharp contrast with the first-principles calculations based on the crystal structure at room temperature, quantum oscillations reveal small Fermi pockets with light effective masses, suggesting significant change in the Fermi surface topology caused by the low temperature structural distortion. No superconductivity emerges in SrAg$_4$As$_2$ down to 2 K under pressures up to 2.13 GPa; instead, the low temperature structural distortion moves up linearly to high temperature at a rate of $\approx$13 K/GPa above 0.89 GPa.
\end{abstract}
\date{\today}
\maketitle
\section{Introduction}

Quasi-low-dimensional materials have attracted immense research interest because they often host exotic ground states due to strong quantum and thermal fluctuations. Layered pnictide materials have provided a rich platform to explore these ground states and the competitions among them. For example, the Fermi surface nesting leads to electronic/magnetic instabilities, resulting in spin-density-wave (SDW) state in Fe pnictide superconductors \cite{NLwang,DaiSDW} and the charge-density-wave (CDW) in RAgSb$_2$ family \cite{canfieldCDW, canfieldCDW1}. For the former, the SDW order competes with the superconductivity and quantum critical behavior was argued in P doped BaFe$_2$As$_2$ family \cite{QCP}. For the latter, the CDW opens a gap at the Dirac point, leading to the exotic properties \cite{HDing}. Among the layered pnictide, the ones containing AgAs have shown rich structural varieties, such as BaAg$_2$As$_2$ \cite{baag2as2} as well the derivatives of the HfCuSi$_2$ structure \cite{doert}. In this paper we focus on the layered ternary pnictide SrAg$_4$As$_2$, which crystalizes in the centrosymmetric trigonal space group $R\bar 3 m$ \cite{SrAg4As2}, as shown in the right inset of Fig. 1(a). It can be taken as ordered SrAg$_2$As$_2$ mixed with itinerant Ag layers, where Ag atoms (light blue balls) in the itinerant Ag layers partially occupy three different sites. As a charge-balanced Zintl phase, first-principles calculations suggest that it is a gapless system \cite{SrAg4As2}. We have performed a systematic study of SrAg$_4$As$_2$ through transport and thermodynamic measurements. We show that at least one non-magnetic phase transition exists in this material and affects the electronic states in the system and results in various anomalies across the transition in its physical properties. In sharp contrast with the first-principle calculations which was performed on the room-temperature crystal structure, by examining the quantum oscillations in this material, we reveal a few much smaller Fermi pockets with light masses, pointing to the change of Fermi surface topology caused by the low temperature structural distortion.

\section{Single crystal growth and experimental methods}

The single crystals of SrAg$_4$As$_2$ were grown via the self-flux method. The starting materials Sr and Ag$_2$As were mixed according to the molar ratio Sr : Ag$_2$As = 1 : 4. They were loaded into a 2-ml alumina crucible and sealed inside an evacuated quartz tube. The growth ampoule was heated to 1100 $^o$C and slowly cooled to 650 $^o$C at a rate of 5 $^o$C/h,  at which temperature the single crystals were separated by decanting the flux in a centrifuge. Sizable plate-like single crystals were harvested, as shown in the left inset of Fig. 1(a). The typical size of the crystals can be up to 3$\times$3$\times$0.5 mm$^3$ and the as-grown surface is the $ab$ plane.

Electrical resistivity, Hall coefficient and heat capacity data up to 300 K were collected using a Quantum Design Physical Property Measurement System (QD PPMS Dynacool). Seebeck coefficient was measured by a homemade probe inserted in the PPMS sample chamber \cite{mun}. The resistivity above 300 K was measured using a homemade probe in a tube furnace, where the tube chamber was connected to a vacuum pump to keep the pressure inside below 20 mTorr. The standard four-probe configuration was used in the electrical resistivity ($\rho_{xx}$) and Hall resistivity ($\rho_{yx}$) measurements. Four Pt wires were attached to the resistivity bar using Dupont silver paste 7713. During the transport measurements, the magnetic field was swept from -9 T to 9 T. The data were then processed to get $\rho_{xx}$ using $\rho_{xx}$(B)=($\rho_{xx}$(B)+$\rho_{xx}$(-B))/2 and $\rho_{yx}$ using $\rho_{yx}$(B)=($\rho_{yx}$(B)-$\rho_{yx}$(-B))/2.
The susceptibility was measured with a QD Magnetic Properties Measurement System (QD MPMS). The high field $\rho_{xx}$ data up to 35 T were collected in high field magnet laboratory (HFML)in Nijmegan.

Pressures up to $\approx$ 2.13 GPa were applied using a Be-Cu piston pressure cell made by C\&T Factory Co. Silicone oil (viscosity 5 cSt (25$^o$C), Sigma-Aldrich) was used as the pressure medium. The hydrostaticity of silicone oil of a wide range of viscosities has been studied and was found to be an excellent pressure medium even up to extremely high pressures \cite{P1,P2}. The feedthrough was sealed using Stycast 2850 FT. We determined the pressure inside the sample space through the pressure dependence of the superconducting transition of Pb.

First-principles calculations based on density functional theory (DFT) were carried out to study the electronic structure. The full-potential linearized augmented plane-wave method and the generalized gradient approximation of the exchange-correlation potential as implemented in the Wien2k package were used \cite{wien2k}.

\section{Experimental results and discussion}

\subsection{Physical properties of SrAg$_4$As$_2$}

\begin{figure}
  \centering
  \includegraphics[width=3.3in]{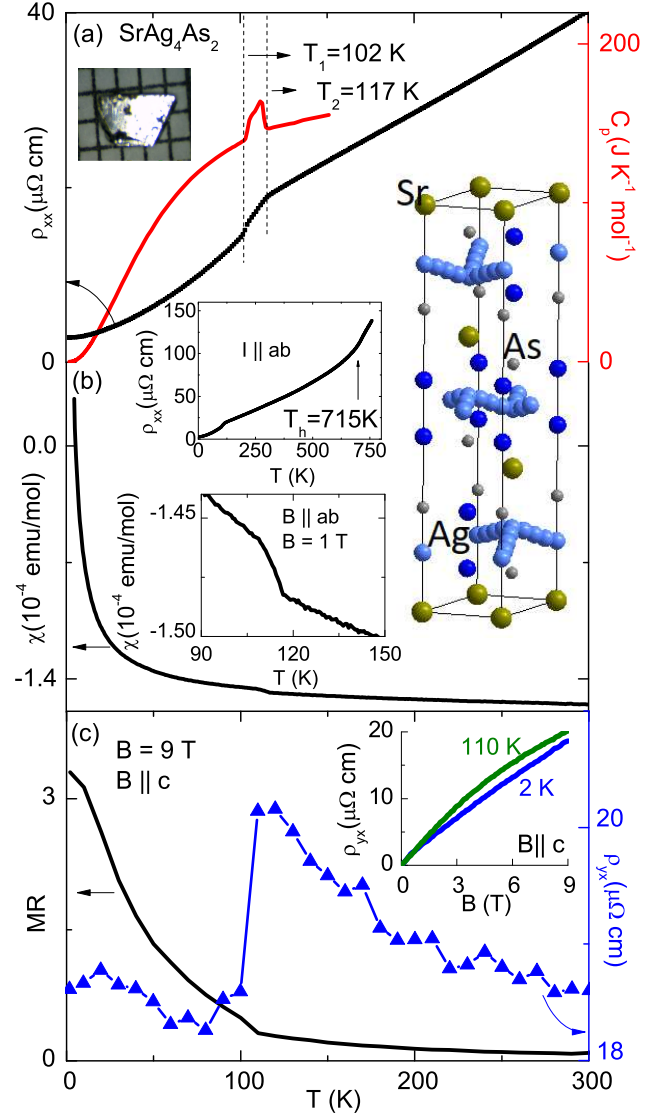}
  \caption{(a): The temperature dependent resistivity and heat capacity. Left inset: A single crystal of SrAg$_4$As$_2$ against a 1-mm scale. Top left inset: The temperature dependent resistivity measured up to 750 K. (b): The temperature dependent susceptibility measured with B $\parallel ab$ at B = 1 T. Bottom left inset: The zoomed-in susceptibility around 110 K. Right inset: The crystal structure of SrAg$_4$As$_2$ at 300 K. Dark green ball: Sr; Dark blue ball: fully occupied Ag sites; Light blue ball: partially occupied Ag sites; Grey ball: As. (c): The temperature dependent Hall resistivity ($\rho_{yx}$) and transverse magnetoresistance (MR) taken at B = 9 T and B $\parallel c$. Inset: The field dependent $\rho_{yx}$ taken with B $\parallel c$ at T = 2 K and 110 K. }
  \label{fig:Fig1}
\end{figure}

The temperature dependent resistivity, heat capacity, susceptibility of SrAg$_4$As$_2$ up to 300 K are presented in Fig. 1(a) and (b). It is a diamagnetic metal with $\rho_0$ around 3$\mu\Omega$-cm. Upon cooling, between T$_1\sim102$ K and $T_2\sim117$ K, besides subsequent slope changes in resistivity as shown in Fig. 1(a) and subsequent slope changes in susceptibility as shown in the inset of Fig. 1 (b), two broad peaks in heat capacity (Fig. 1(a)), which are distinct from the heat capacity with a first-order phase transition \cite{first}, are observed, featuring second order phase transitions. The Debye temperature inferred from the low temperature heat capacity is 140 K. The diamagnetic susceptibility is mainly due to the core diamagnetism of the heavy silver atoms and Landau diamagnetism, suggesting that the phase transitions lack a magnetic nature.
Hereafter, we will call these transitions as the low temperature phase transitions. The left inset of Fig. 1(a) shows the temperature dependent resistivity data up to 750 K. Upon warming, besides the slope changes between 102 K and 117 K, an additional resistivity slope change was observed at $T_h \sim$ 715 K, suggesting an additional high temperature non-magnetic phase transition in this compound.

The temperature dependent Hall resistivity $\rho_{yx}$ and MR taken at 9 T are shown in Fig. 1(c). While the non-linear field dependence of $\rho_{yx}$ shown in the inset of Fig. 1(c) suggests multi-band transport, the positive sign of $\rho_{yx}$ suggests that holes dominate the transport. Anomalies appear in both $\rho_{yx}$(T) and MR(T). Upon cooling, $\rho_{yx}$ first slightly increases by 8\% from 300 K to the transition temperature, where it exhibits a sudden drop back to the room-temperature value; it then roughly remains constant at lower temperatures. In contrast to the roughly temperature independent $\rho_{yx}$, upon cooling, the magnitude of the MR first increases by 50\% across the transition; then it monotonically grows by 7 times down to 2 K. A sharp increase of the MR in the low temperature region was observed in various nonmagnetic multiband systems, such as the type-II Weyl semimetal WTe$_2$, the weak topological insulator NbAs$_2$, the Weyl semimetal TaAs, etc. \cite{WTe2,NbAs2,TaAs1,TaAs2}, where the compensation of holes and electrons can induce the resistivity upturn (so called turn-on effect) at low temperatures \cite{kwok}. For SrAg$_4$As$_2$, the 50\% increase in MR likely arises due to the phase transition while the further growth of the MR at lower temperatures may originate from the turn-on effect.

Non-magnetic phase transitions can be electron-driven or lattice-driven. Electron-driven Mott phase transition is unlikely here since strong Coulomb interaction is unexpected for Ag$^{1+}$ ions. Lattice-driven structural phase transition is a probable candidate. Meanwhile, considering that Fermi surface nesting may show up in materials with layered structure, electron-driven charge density wave (CDW) formation is a possible candidate too.
 The chance of hosting a CDW transition may not be high since the resistivity anomaly in SrAg$_4$As$_2$ does not show the hump feature of a typical CDW phase transition. However, because the CDW may only gap a small portion of the Fermi surface and the formation of CDW can significantly increase the scattering rate, we still can not exclude the possibility. Further investigation using scanning tunneling microscope is needed to check if electron density modulation exists so that the two scenarios can be distinguished. Nevertheless, in both cases, structural distortions will take place. Therefore, hereafter, we will call the transitions around 110 K as low temperature structural distortions.

\subsection{Magnetotransport properties of SrAg$_4$As$_2$}

Angle-dependent transverse MR (AMR) measured with the field B perpendicular to the current is a powerful tool to probe the Fermi surface topology since the AMR is sensitive to the mobility tensor $\hat \mu  =\tau \hat m^{-1}$, where $\tau$ is the relaxation time and $\hat m$ is the mass tensor closely related to the Fermi surface topology \cite{Bi}. Figures 2(a) and (b) show the AMR taken at 9 T at various temperatures with B rotating in the planes perpendicular to the $c$ axis and $a$ axis, respectively. The AMR in Fig. 2(a) with B $\perp c$ shows the threefold angular oscillations with very little anisotropy at all temperatures, consistent with the threefold rotational symmetry of the $ab$ plane and suggesting small in-plane anisotropy. In sharp contrast, the AMR in Fig. 2(b) with B $\perp a$ shows a twofold rotational symmetry, where strong anisotropy with a ratio between the maxima and minima of about 3.5 exists, suggesting a quasi-two-dimensional electronic structure in this compound. The AMR reaches its minimum value when B $\parallel ab$ plane (0$^\circ$) and gradually increases when the field rotates toward the $c$ axis (90$^\circ$) with the maxima somewhere around 45$^\circ$. This morphology suggests a more complex Fermi surface than a simple two-dimensional one. We noticed that at temperatures higher than 100 K, the feature at 45$^\circ$ in Fig. 2(b) is gone. It raises the question of whether the appearance of this feature is related to a significant Fermi surface reconstruction caused by the low temperature structural distortion. To clarify this, we measured AMR at 2 K with B $\perp a$ under various fields, as shown in Fig. 2(c). The solid line is the simulated curve based on a cylindrical Fermi surface assumption. At lower field, the AMR agrees well with the simulated curve and the maximum AMR appears at B $\parallel c$. With increasing field, deviations appear and the direction where the maximum AMR occurs move further and further away from the $c$ axis. Therefore, the deviation from the two-dimensional Fermi surface observed in Figures 2(b)-(c) is more likely to be due to the fact that the mobility tensor of carriers associated with each Fermi pocket have different temperature and field dependencies.

\begin{figure}
  \centering
  \includegraphics[width=3.5in]{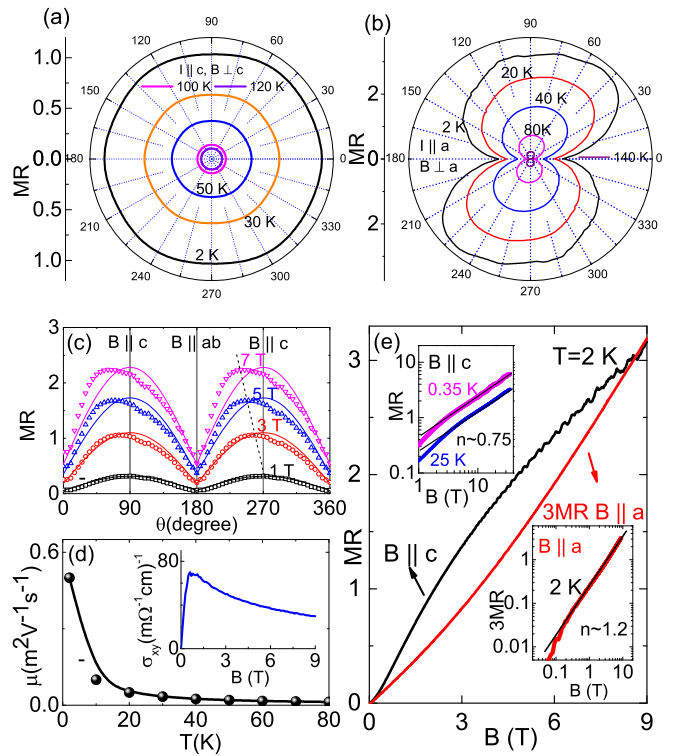}
  \caption{(a)-(b): The polar plot of the transverse angle-dependent magnetoresistance (AMR) of SrAg$_4$As$_2$ at various temperatures with B $\perp c$ and B $ \perp a$ at 9 T, respectively. B is perpendicular to the current flow. (c) The AMR of SrAg$_4$As$_2$ at various fields at 2 K. Hollow symbols: Experimental data. Solid lines: Simulated AMR assuming cylindrical Fermi pocket. (d): The temperature dependent average mobility. Inset: The field dependent electrical conductivity $\sigma_{xy}$ at 2 K. (e) The field dependent MR (MR(B)) at 2 K with B $\parallel c$ or $\parallel a$ up to 9 T. Top inset: The log-log plot of the MR(B) at 2 K with B $\parallel c$ up to 35 T. Bottom inset: The log-log plot of the MR(B) at 2 K with B $\parallel a$.
  }
  \label{fig:Fig3}
\end{figure}

Figure 2(d) shows the temperature dependent geometric mean of the mobilities $\bar \mu$ while the inset shows the field dependent electrical conductivity $\sigma_{xy}$ at 2 K which was calculated using $\sigma_{xy}=\rho_{yx}/(\rho_{xx}^2+\rho_{yx}^2)$. According to the standard Bloch-Boltzmann transport theory \cite{Cd3As2}, we inferred the geometric mean of the mobilities $\bar \mu\equiv\sqrt{\mu_1\mu_2}$ using the formula $\bar \mu$=1/B$_{max}$, where B$_{max}$ is the field at which $\sigma_{xy}$(B) peaks at. $\bar \mu$ increases with decreasing temperature with the value of 0.5 m$^2$V$^{-1}$s$^{-1}$ at 2 K. This value is lager than that in the nodal line semimetal candidate CaAgAs \cite{CaAgAs} but smaller than those in WTe$_2$, Cd$_3$As$_2$, etc. \cite{WTe2,Cd3As2}.

Figure 2(e) shows the field dependent MR up to 9 T with the current along the $a$ axis. With B $\parallel a$, MR shows convex behavior with field. Instead of the quadratic field dependence, MR is roughly proportional to B$^{1.2}$ (the bottom inset of Fig. 2(e)). With B $\parallel c$, MR shows concave behavior with field, which may be a trend of saturation. However, even with fields up to 35 T along the $c$ axis, no saturation is realized up to 25 K, the highest temperature we measured. MR is roughly proportional to B$^{0.75}$ in an extended field range from 2 T to 35 T and reaches 600\% at 0.35 K (the top inset of Fig. 2(e)). This field dependence holds from 3 T to 35 T at 25 K. Non-quadratic field dependence of MR is non-trivial and has been observed in many materials  such as the Dirac semimetal Cd$_3$As$_2$ , the Weyl semimetal TaAs etc. \cite{Cd3As2,TaAs1,TaAs2}.
The origin of this non-quadratic field dependence can arise from distinct mechanisms. Magnetism can be excluded since SrAg$_4$As$_2$ is nonmagnetic \cite{SrAg4As2}. In the classical region, an open Fermi surface can lead to linear MR and a saturating Hall coefficient. This is not the case here either \cite{AB Pippard}. In the quantum limit, MR will be linear when only the lowest Landau band participates in the conductivity \cite{Burkov,Goswami}. However, the non-quadratic response in SrAg$_4$As$_2$ appears at very low field region which is away from the quantum region. Disordered systems with high mobilities, such as Ag$_2$Se, Dirac semimetal Cd$_3$As$_2$ and few-layer black phosphorus \cite{Littlewood,Cd3As2,coldea,P} also show linear MR. This has been investigated using the classical disorder model \cite{coldea}, where Monte Carlo simulation shows that the linear MR in disordered materials with high mobility can come from the spatial fluctuation of local current density. As we will show later, indeed, the charge carriers in SrAg$_4$As$_2$ have high mobilities and small effective masses. Therefore, the linear MR in SrAg$_4$As$_2$ may suggest that local current density fluctuation plays a key role here.

\begin{figure*}
  \centering
  \includegraphics[width=6in]{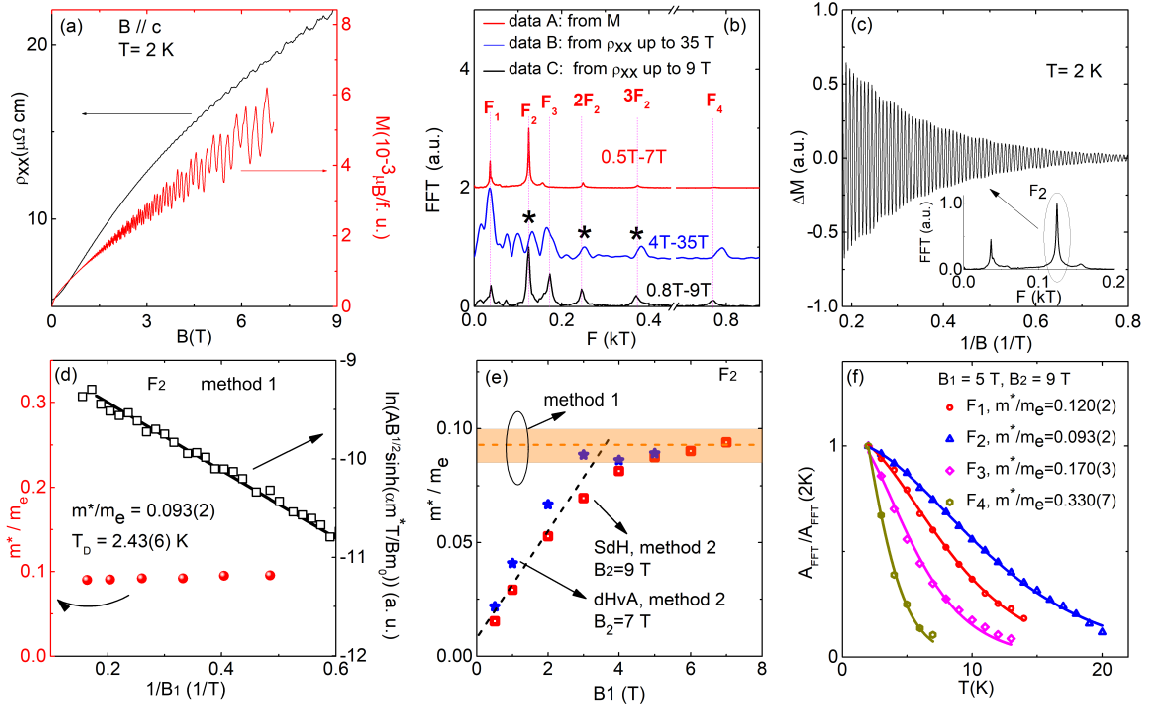}
  \caption{(a): The field dependent resistivity and magnetization at 2 K with B $\parallel c$.  (b): The FFT spectra obtained from $M_{osc}$ (data A), HFML $\rho_{xx_{osc}}$ (data B) and PPMS $\rho_{xx_{osc}}$ (data C). Polynomial fitting is used to remove the smooth background from $M(B)$ and $\rho_{xx}$ to get $M_{osc}$ and $\rho_{xx_{osc}}$. The FFT was performed from 0.5 T to 7 T on $M_{osc}$ (data A), from 4 T to 35 T on HFML $\rho_{xx_{osc}}$ (data B), and  from 0.8 T to 9 T on PPMS $\rho_{xx_{osc}}$ (data C). The fundamental oscillation peaks and the harmonic peaks are labeled as F$_1$, F$_2$, 2F$_2$, 3F$_2$, F$_3$ and F$_4$. (c): $M_{osc}$ associated with F$_2$ (data A) vs. $1/B$ at 2 K. The method to obtain $M_{osc}$ is described in the text. Inset: the detail of the FFT spectrum near $F_2$. (d) The Dingle plot and the evolution of the cyclotron effective mass of F$_2$ with $B_1$. Method 1 as described in the text is used for the data analysis. (e) The evolution of the cyclotron effective mass of F$_2$ with $B_1$. Method 2 as described in the text is used for the data analysis. The orange rectangle shows the cyclotron effective mass of F$_2$ obtained using method 1 with 99\% of confidence level (2.576 times the standard error).
  (f) The normalized $A^{FFT}/A^{FFT}(2 K)$ of F$_1$, F$_2$, F$_3$ and F$_4$. The FFT was performed on PPMS $\rho_{xx_{osc}}$ (data C) from 5 T to 9 T. Fits by Lifshits-Kosevich formula are denoted by solid lines.}
  \label{fig:Fig3}
\end{figure*}

\subsection{Quantum oscillations of SrAg$_4$As$_2$}

Quantum oscillations are observed in both resistivity and magnetization measurements above 1 T at low temperatures, as shown in Fig. 3(a). To shed light on the Fermi surface topology in the low temperature structure, we performed a systematic study of Shubnikov-de Haas (SdH) quantum oscillations (QO) and de Haas-van Alphen (dHvA) QO for SrAg$_4$As$_2$. According to the Onsager relation, the area $S$ of the extreme cross section of the Fermi surface which is perpendicular to the applied field are related to the oscillation frequency $f$ by the equation $f=\frac{\hbar}{ 2\pi}S$ \cite{Onsager relation}.
Figure 3(b) presents the Fast Fourier transformation (FFT) spectra of the oscillating $\rho_{xx_{osc}}$ and $M_{osc}$, where $\rho_{xx_{osc}}$ and $M_{osc}$ are obtained by subtracting a polynomial background from $\rho_{xx}$ and $M$, respectively. Different oscillation frequencies are detected. Combining all three sets of data measured on different samples (see Supplemental Material \cite{supp}), we were able to assign the oscillations to four fundamental frequencies ($F_1$ = 38 T, $F_2$ = 125 T, $F_3$ = 173 T, $F_4$ = 760 T) and the harmonics of F$_2$, as labeled in Fig. 3(b).

Quantum oscillations are a function of both $B$ and $T$. For the dHvA oscillations which occur in the magnetization, $M_{osc}(B, T)$ can be described by the Lifshits-Kosevich (LK) formula \cite{Onsager relation}
\begin{equation}
M_{osc}(B, T)\propto (B)^{\lambda}R_TR_DSin[2\pi(F/B+\gamma-\delta)]
\end{equation}
\begin{equation}
A(B, T)\propto (B)^{\lambda}R_TR_D
\end{equation}
\begin{equation}
R_T=\frac{\alpha Tm^*/B m_e}{sinh(\alpha Tm^*/B m_e)}
\end{equation}
\begin{equation}
R_D=exp{(-\alpha T_Dm^*/B m_e)}
\end{equation}
where $A(B, T)$ is the amplitude of the QO, $R_T$ is the thermal damping term which dominate the temperature dependence of the oscillations, $R_D$ is the Dingle damping factor which dominates the field dependence of the oscillations, $F$ is the oscillation frequency, $T$ is the temperature, $m_e$ is the free electron mass, $\alpha=2\pi^2k_Bm_e/e\hbar=14.69 $ T/K, $m^*$ is the cyclotron effective mass, and $T_D=\frac{\hbar}{2\pi k_B\tau_s}$ is the Dingle temperature which is a measure of the single-particle scattering rate $\tau_s$.
$\lambda$ is a dimensional factor, which is 1/2 for the three-dimensional (3D) case
and 0 for the two-dimensional (2D) case. The phase factor $\gamma-\delta$ and the Berry phase $\phi_B$ are related by the equation $\phi_B=\pi-2\pi\gamma$ and the phase shift $\delta$ is 0 or $\pm$1/8 for 2D or 3D Fermi surfaces respectively.

Let us take the dHvA oscillation as an example to explain how data analysis of QO is performed. The experimental QO of a material are a summation of individual oscillations associated with each oscillation frequency. If only one quantum oscillation frequency exists, the data analysis to infer $m^*$ and $T_D$ is straightforward and is described as follows \cite{caagas} (method 1). Firstly, by subtracting a smooth background of $M(B)$, we can obtain the oscillation part $M_{osc}(B)$. Then at each temperature, we extract the amplitude of the $M_{osc}(B)$ at a fixed $B_1$ so that a plot of the temperature dependent oscillation amplitude, $A(T)$, can be made.
By fitting $A(T)$ using Eq. (1) with $B=B_1$, the $m^*$ can be obtained without approximation. Then at a fixed temperature $T_1$, we can make a plot of the field dependent oscillation amplitude, $A(B)$. By fitting $A(B)$ using Eq. (1) with $T=T_1$ and $m^*$, the Dingle temperature $T_D$ can be determined. Finally, using $m^*$ and $T_D$, and by fitting the $M_{osc}(B)$ at the fixed temperature $T_1$, the Berry phase can be obtained.

However, in most cases, more than one oscillation frequencies exist. If we do a Fast Fourier Transformation (FFT) of $M_{osc}(B)$ and these frequencies are well separated, we can filter out the individual oscillations associated with each frequency \cite{NbAs2}. A sanity check can be made to see if the summation of these individual oscillations agree with the total experimental oscillation. Then for each individual oscillatory pattern, method 1 can be used to infer $m^*$ and $T_D$ for that frequency/Fermi pocket without approximation. With the known $m^*$s and $T_D$s, a multi-band LK fitting of $M_{osc}(B)$ at a fixed temperature $T_1$ can be made using Eq. (1) to determine the Berry phase associated with each frequency.

\begin{figure*}
  \centering
  \includegraphics[width=6.5in]{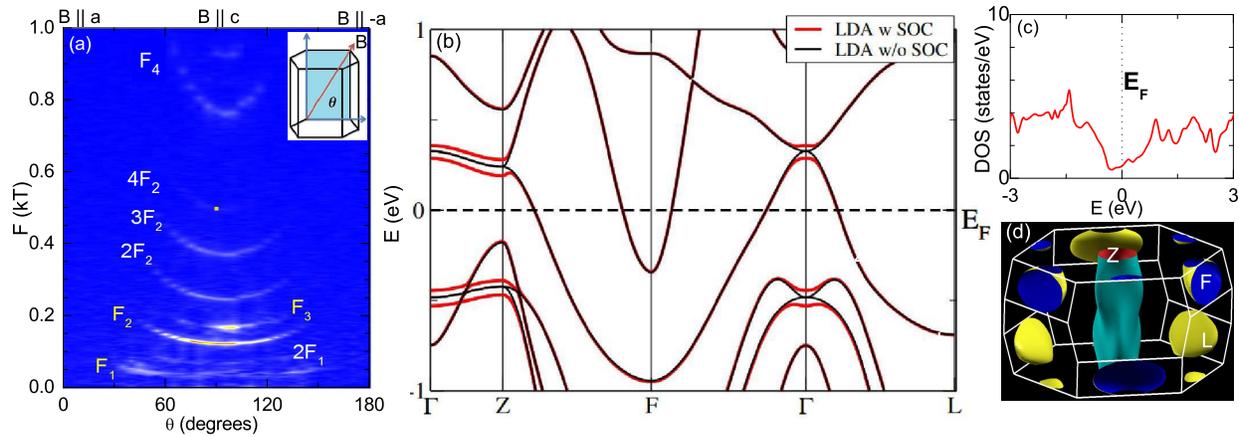}
  \caption{(a): The color plot of angle-dependent FFT of $\rho_{xx_{osc}}$ with the rotation geometry shown in the inset. (b): The band structure calculated from first-principles calculations with density-functional theory (DFT) using the crystal structure at 300 K. (c): The density of states based on the DFT calculation. (d): The Fermi surface based on the DFT calculation using the crystal structure at 300 K.}
  \label{fig:Fig3}
\end{figure*}

For the above two situations, no approximation of the LK formula is needed during the data analysis. Now, let us move to the most complicated situation where those frequencies are not well separated but overlap with each other and  as a result, individual oscillations cannot be obtained without beats. In this situation, standard FFT data analysis method with an ``average" $B$ has to be used (method 2). This is the case for SrAg$_4$As$_2$.
Firstly, we performed FFT on $M_{osc}(B)$ in the FFT window of $B_1$ and $B_2$ at various temperatures. Secondly, we collected the amplitude of the FFT, $A^{FFT}$, for each QO FFT peak at various temperatures to get $A^{FFT}(B, T)$. We then fit the data with Eq. (1) to obtain $m^*$ where an ``average" $B$ of $2/(1/B_1+1/B_2)$ is used. 

For SrAg$_4$As$_2$, we found that F$_2$ in the dHvA oscillations can be relatively nicely band-pass filtered into oscillations with only weak beats, as shown in Fig. 3(c). We then used method 1 to analyze the pattern in Fig. 3(c) to get $m^*$. Six different values were chosen for $B_1$, and the resulted $m^*$ are shown as red spheres in Fig. 3(d). They show very small variation, with $m^*/m_e=0.093(2)$. Using this value for $m^*$, $T_D$ was found to be 2.43(6) K, suggesting a long quantum life time of $\tau_s=0.50(1)$ ps. This value is a few times longer than that of Cd$_3$As$_2$ but similar to the one of ZrTe$_5$\cite{coldea, zrte5}. Since SrAg$_4$As$_2$ shows partial occupancy which is a necessary condition for the chemical bonding rules to be satisfied \cite{SrAg4As2}, it is surprising to see the charge carriers can have such a long quantum life time.

We then obtained $m^*$ of $F_2$ using method 2 by varying $B_1$ from 0.5 T to 7 T, as shown in Fig. 3(e). We found that $m^*$ varies much with the choice of $B_1$. In Fig. 3(e), we also plotted $m^*$ of F$_2$ obtained using method 1 with 99\% of confidence level (2.58 of standard error) as shown in the orange rectangle. We can see that as we increase $B_1$ from 0.5 T to 7 T, the $m^*$ first increases sharply and then starts saturating when $B_1$ $>$ 4 T. The saturated value is close to the $m^*$ we obtained using method 1. This observation suggests that if a large FFT window is chosen, the effect of the $R_D$ term may lead to an unreal effective mass using method 2. Therefore, based on the above observation and considering the fact that a sufficient number of oscillations are necessary for a good FFT analysis, we chose $B_1$ as 5 T and $B_2$ as 9 T to determine the $m^*$ for all other frequencies. The LK fit of the temperature dependent FFT amplitude gives a fitting error around 1\% for all frequencies. Thus, to be more conservative, we used a standard error of 2\% for all cyclotron masses, which is the one obtained for $F_2$ using method 1. The resulting effective masses are 0.120(2) for F$_1$, 0.093(2) for F$_2$, 0.170(3) for F$_3$ and 0.330(7) for F$_4$.

To further investigate the Fermi surface topology, we studied the angle-dependent SdH oscillations and compared them with the first-principles calculations. The measurement geometry is shown in the inset of Fig. 4(a). $B$ was rotated from the $a$ axis to the $c$ axis and then to the -$a$ axis. The angle-dependent FFT spectra of $\rho_{xx_{osc}}$ obtained using data C are presented in Fig. 4(a). Both F$_2$ and F$_3$ have minima at B $\parallel c$ and increase with the field B rotating away from the $c$ axis, suggesting that the associated Fermi pockets have the minimum extreme cross sections in the plane perpendicular to $c$.

To understand the observed SdH, DFT band structure calculations were performed using the crystal structure in Ref. \cite{SrAg4As2} obtained at 300 K. The referenced crystal structure shows that there are four Ag sites. Two of them have very low occupancies (0.051(3) and 0.036(4)), one has a higher occupancy of 0.789(8) and another is set to be fully occupied. To simplify the calculation, the Ag sites with low partial occupancies are ignored while the Ag site with larger partial occupancy is fully occupied. Figure 4(c) shows the density of states (DOS) plot. It suggests a semimetallic character for SrAg$_4$As$_2$, consistent with experiments. The DOS at the Fermi level is dominated by arsenic $p$ orbitals while the Ag $d$ orbital weights small. Therefore, although our DFT calculation employed simplified crystal structure ignoring the Ag sites with low occupancies, it should keep the main features of the Fermi surface. Figure 4(b) shows the DFT band structure. The feature of the band structure near the Fermi level is not affected by spin-orbit coupling. Two types of Fermi pockets exist, as shown in Fig. 4(d). The hole pocket centered at the $\Gamma$ point is two dimensional along the $\Gamma$-Z direction with the effective mass of 0.4$m_e$ when B $\parallel c$ while the electron pocket centered at the F point is a three dimensional oblate ellipsoid with the effective mass of 0.5$m_e$ when B $\parallel c$, consistent with the electron-hole compensation suggested by the observed turn-on effect. However, the inferred frequencies from the minimum extreme cross section of the two Fermi pockets are around 1000 T, much larger than F$_1$ to F$_3$ (7 T, 125 T and 173 T) we observed experimentally but reasonably close to the F$_4\sim 780 $ T.

Considering the small SdH Fermi pockets with extremely small effective masses and the dramatic discrepancy between QO and DFT calculations, the Fermi surface topology must have changed significantly due to the low temperature structural distortion. Unexpected small Fermi pockets have been discovered in Cuprates, as the result of the band folding effect due to the formation of CDW \cite{YBCOCDW, YBCOCDW1} while banding folding has been observed in Fe-based superconductors due to the structural and magnetic phase transitions \cite{james}. At low temperatures, the structural modulation expands the unit cell and lowers the symmetry of the crystal. As a consequence, the Brillouin zone is folded and small Fermi pockets can appear. Further investigation is needed to clarify the origin of this low temperature structural distortion and thus to explain if the small Fermi pockets observed are from a simple band folding effect.

\subsection{The effect of external pressure on SrAg$_4$As$_2$}

Tuning the phase transitions using either chemical doping/intercalation or external pressure may result in unusual ground states. For example, suppressing the CDW order in TiSe$_2$ using Cu intercalation has led to the appearance of superconductivity \cite{CuTiSe2}. The suppression of SDW in LaFeAsO by F doping on the O sites has made the discovery of the second high temperature superconducting family \cite{DaiSDW}. To tune the ground state of SrAg$_4$As$_2$, we applied external pressure using silicone oil (viscosity 5 cSt (25$^o$C), Sigma-Aldrich) as the pressure medium on a piece of SrAg$_4$As$_2$ whose RRR is $\approx$18 at ambient pressure. Figure 5(a) shows the evolution of $\rho$(T) under the application of external pressure up to 2.13 GPa. No superconductivity is observed. Figure 5(b) shows the first derivative of the $\rho$(T) at various pressures. Below 0.89 GPa, the transition temperature is slightly suppressed with pressure at a rate of 3.2 K/GPa and then starts to increase linearly at a rate of $\approx$ 13 K/GPa up to 2.13 GPa, the highest pressure we applied. The T-P phase diagram is summarized in Fig. 5(c) where the non-magnetic transition temperature $T_s$ is determined by the criterion shown in Fig. 5(b). The T-P phase diagram shows a v-shaped pressure dependence of the transition temperature. It is worth noting that silicone oil solidifies at around 1 GPa at room temperature, above which the hydrostaticity is lost. This is evidenced by the peak broadening above 0.89 GPa in Fig. 5(b). However, we believe the enhancement of the $T_s$ above 0.89 GPa is intrinsic since the $T_s$ increased by $\sim 20$ K from 0.89 GPa to 2.2 GPa and shows no sign of slowing down. The drop in resistivity across the transition, $\Delta \rho_{xx}$, was determined using the criterion shown in the inset of Fig. 5(a). $\Delta \rho_{xx}$ first decreases and then increases with pressure, in a similar manner as $T_s$, as shown in Fig. 5(c). Both pressure dependent transition temperature and $\Delta \rho_{xx}$ that we observe here are reminiscent of the evolution of the CDW transition in SmNiC$_2$ under pressure \cite{smnic2-1, smnic2-2}.

\begin{figure}
  \centering
  \includegraphics[width=3.5in]{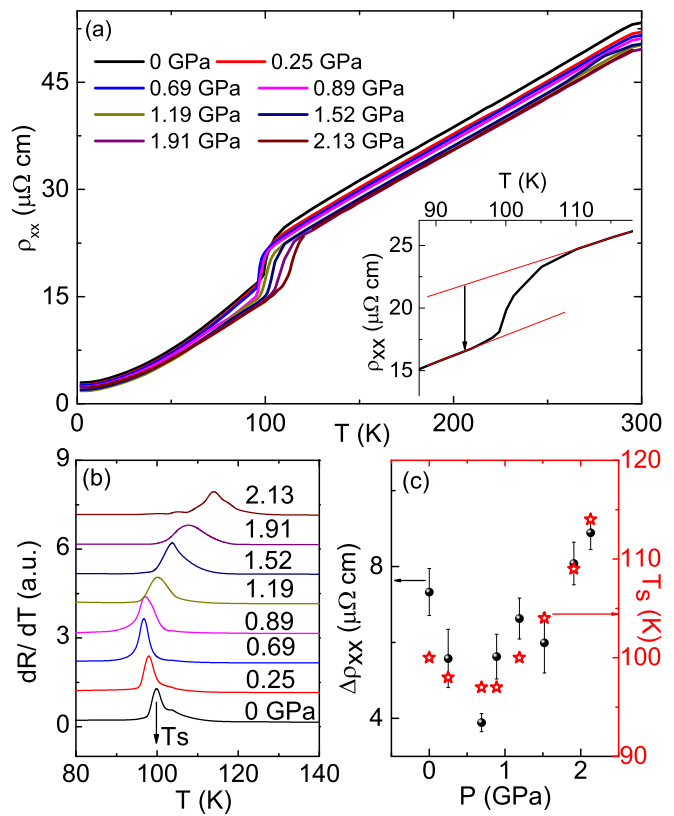}
  \caption{(a): The temperature dependent resistivity of SrAg$_4$As$_2$ under various pressure. Inset: The criterion to determine the $\Delta\rho_{xx}$. (b) The $d$R/$d$T at various pressure. The criterion to determine T$_s$ is shown. (c) The temperature-pressure (T-P) phase diagram and the drop in resistivity $\Delta \rho_{xx}$ across the transition as a function of pressure. The error bars correspond to the standard deviation of an average of six calculations.}
  \label{fig:Fig3}
\end{figure}

\section{Conclusion}
In conclusion, SrAg$_2$As$_2$ shows electronic changes across the low temperature structural distortion around 110 K. In addition, a high temperature structural distortion around 715 K was evidenced by a slope change in resistivity. Angle-dependent MR measurements show small in-plane anisotropy but large out-of-plane anisotropy, suggesting a two-dimensional electronic structure. Multi-band transport is observed; large non-saturating MR was observed with high mobility, consistent with the semimetallic nature revealed by the DFT calculation. MR shows nontrivial field dependence, which may be due to charge fluctuation in this highly disordered material. Both dHvA and SdH oscillation reveal small Fermi pockets with light effective masses. An unusually long quantum life time of 0.50(1) ps was determined for the Fermi pocket associated with frequency F$_2$. However, the DFT calculation based on the room temperature crystal structure leads to Fermi pockets with much larger size and heavier masses. We attribute this discrepancy to the significant change in the Fermi surface topology caused by the low temperature structural distortion. The application of external hydrostatic pressure first suppresses the transition below 0.89 GPa and then enhances the phase transition linearly at a rate of $\approx$13 K/GPa up to 2.13 GPa without inducing superconductivity.

\section*{Acknowledgments}
Work at UCLA was supported by the U.S. Department of Energy (DOE), Office of Science, Office of Basic Energy Sciences under Award Number DE-SC0011978. AIC acknowledges an EPSRC Career Acceleration Fellowship (EP/I004475/1). Part of this work was supported by HFML-RU/FOM and LNCMI-CNRS, members of the European Magnetic Field Laboratory (EMFL) and by EPSRC (UK) via its membership to the EMFL (grant no. EP/N01085X/1). Work at Rutgers was supported by the NSF DMREF program under the award NSF DMREF project DMR-1435918. NN would like to thank useful discussion with Gui Xin and Prof. Weiwei Xie.

\end{document}